\documentclass[final,5p,times,twocolumn]{elsarticle}
\usepackage{lineno,hyperref,amsmath,amsthm,amssymb,amsfonts,ragged2e,color,subfig}
\begin{document}
\begin{frontmatter}
\title{Modulational instability of dust-ion-acoustic waves and associated first and second-order rogue waves in super-thermal plasma}
\author{B.E. Sharmin$^{*, 1}$, R.K. Shikha$^{**, 1}$, N.K. Tamanna$^{***, 1}$, N.A. Chowdhury$^{\dag, 2}$,
A. Mannan$^{\ddag, 1}$, and A.A. Mamun$^{\S, 1}$}
\address{$^{1}$Department of Physics, Jahangirnagar University, Savar, Dhaka-1342, Bangladesh\\
$^{2}$Plasma Physics Division, Atomic Energy Centre, Dhaka-1000, Bangladesh\\
e-mail: $^*$sharmin114phy@gmail.com, $^{**}$shikha261phy@gmail.com, $^{***}$tamanna1995phy@gmail.com,\\
$^{\dag}$nurealam1743phy@gmail.com, $^{\ddag}$abdulmannan@juniv.edu, $^{\S}$mamun\_phys@juniv.edu}
\begin{abstract}
A proper theoretical research has been carried out to explore the modulational instability (MI) conditions
of dust-ion-acoustic (DIA) waves (DIAWs) in a three-component dusty plasma system containing inertialess
$\kappa$-distributed electrons, and inertial warm positive ions and negative dust grains. The
novel nonlinear Schr\"{o}dinger equation (NLSE) has been derived by employing reductive perturbation method.
The analysis under consideration demonstrates two types of modes, namely, fast and slow DIA modes.
The dispersion and nonlinear properties of the plasma medium as well as the MI conditions of DIAWs and the
configuration of the  energetic rogue waves (RWs) associated with DIAWs in the Modulationally unstable regime have been rigorously changed by the
plasma parameters, namely, charge, mass, temperature, and number density of the plasma species. The findings of our
investigation will be useful in understanding the criteria for the formation of electrostatic RWs in both
astrophysical environments (viz., Jupiter's magnetosphere, cometary tails, Earth's mesosphere, Saturn's rings, etc.)
and laboratory experiments (viz., Q-machines and Coulomb-crystal).
\end{abstract}
\end{frontmatter}
\section{Introduction}
\label{1sec:Introduction}
The massive dust grains have been observed in astro-physical plasmas, viz.,
Jupiter's magnetosphere \cite{Horanyi1993,Horanyi1996},
cometary tails \cite{Horanyi1996}, Earth's mesosphere \cite{Havnes1996}, Saturn's rings \cite{Havnes1995}, and also
laboratory plasmas, viz., Q-machines \cite{Kim2006},  Coulomb-crystal \cite{Chu1994,Thomas1994,Zheng1995}.
Generally, the dust grain is million to billion times heavier than the ion, and the charges which are living onto the
dust grain are thousand times more than ion \cite{Shukla2002}.
The size, mass, and charge of the dust grains are considered to be responsible to change the dynamics
and the criteria for the formation of nonlinear electrostatic waves \cite{Barkan1995,Barkan1996,Shukla1992}.
Barkan \textit{et al.} \cite{Barkan1996} experimentally observed that the phase speed of the ion-acoustic waves increases in
the presence of negatively charged dust grains.

The highly energetic particles exhibit the deviation from the thermally equilibrium state, and can be identified in
Saturn's magnetosphere \cite{Armstrong1983,Barbosa1993,Schippers2008,Young2005}, Earth's bow shock \cite{Asbridge1968},
the solar wind \cite{Marsch1982}, and laboratory experiments \cite{Hellberg2000}. These highly energetic particles are
governed by super-thermal ($\kappa$)-distribution \cite{Vasyliunas1968}, and the parameter $\kappa$ in the $\kappa$-distribution
presents the deviation of energetic particles from the thermally equilibrium state.
The $\kappa$-distribution becomes Maxwellian distribution for large values of $\kappa$ (i.e., $\kappa\rightarrow\infty$) \cite{Chowdhury2019,Emamuddin2018,Atteya2018,Saini2016}. Emamuddin and Mamun \cite{Emamuddin2018} examined the
dust-acoustic shock waves in a multi-component plasma medium having super-thermal electrons, and
found that the width of the shock profile decreases with the super-thermal effect of the electrons.
Atteya \textit{et al.} \cite{Atteya2018} examined the dust-ion-acoustic (DIA) solitary waves in a super-thermal palsma,
and reported that the amplitude and width of the DIA solitary waves increase with decreasing the super-thermality of the
plasma species. Saini and Sethi \cite{Saini2016} analysed the characteristics of DIA cnoidal waves in the presence of
super-thermal electrons.

The nonlinear Schr\"{o}dinger equation (NLSE) is considered as one of the most useful equations for
describing the modulational instability (MI) conditions of different kinds of waves and associated
energy re-localization \cite{Akhmediev2009,Akhmediev2009a,Guo2013}. Rogue waves (RWs), which are the
rational solution of the NLSE \cite{Akhmediev2009,Akhmediev2009a,Guo2013},
can be observed  in super-fluid helium \cite{Ganshin2008}, stock-market crashes \cite{Zhen-Ya2010},
optics \cite{Solli2007,Dudley2019}, and plasma \cite{Tsai2016,Kourakis2003,Javidan2014,Shalini2015}.
Kourakis and Shukla \cite{Kourakis2003} studied the MI and localized excitations of DIA Waves (DIAWs).
Javidan and Pakzad \cite{Javidan2014} considered a three-component plasma system having cold inertial
ions, super-thermal $\kappa$-distributed electrons, and immobile negative dust grains, and studied
the MI of DIAWs by using NLSE, and found that the critical wave number for which the DIAWs becomes
modulationally unstable decreases  with $\kappa$. Shalini and Saini \cite{Shalini2015} investigated
on DIA RWs (DIARWs) in a three-component plasma containing inertial warm ion, inertialess $\kappa$-distributed
electrons, and stationary dust grains, and highlighted that the amplitude of the DIARWs increases with
increasing $\kappa$. To the best knowledge of the authors, no attempt has been made to study the MI
of the DIAWs and associted DIARWs in a three-component plasma having inertial warm positive ion and
negative dust grain, and inertialess $\kappa$-distributed electrons. The aim of the present investigation
is, therefore, to develop NLSE and investigate DIARWs in a three-component dusty plasma.

The arrangement of the paper is as follows: The basic equations are represented in section \ref{1sec:Governing Equations}.
The derivation of NLSE is demonstrated in section \ref{1sec:Derivation of the NLSE}. The MI of DIAWs and rogue waves are provided
in section \ref{1sec:Modulational instability and rogue waves}. The results and discussion are manifested in section \ref{1sec:Results and discussion}.
The conclusion is given in section \ref{1sec:Conclusion}.
\section{Governing equations}
\label{1sec:Governing Equations}
We consider an unmagnetized dusty plasma system consisting of inertial warm positive
ions (i.e., mass $m_i$; charge $Z_i$; temperature $T_i$) and
negative dust grains (i.e., mass $m_d$; charge $Z_d$; temperature $T_d$), and inertialess
super-thermal electrons (i.e., mass $m_e$; charge $e$; temperature $T_e$). The charge neutrality condition
at equilibrium for our considered plasma system can be written as $n_{e0}+Z_dn_{d0}\approx Z_i n_{i0}$;
where $n_{e0}$, $n_{d0}$, and $n_{i0}$ are the equilibrium number densities of super-thermal electrons,
negative dust grains, and positive ions, respectively. The normalized governing equations for our plasma model can be written as
\begin{eqnarray}
&&\hspace*{-1.3cm}\frac{\partial n_d}{\partial t}+\frac{\partial}{\partial x}(n_d u_d)=0,
\label{1eq:1}\\
&&\hspace*{-1.3cm}\frac{\partial u_d}{\partial t}+ u_d\frac{\partial u_d}{\partial x}+\alpha_1n_d\frac{\partial n_d}{\partial x}=\alpha_2\frac{\partial\phi}{\partial x},
\label{1eq:2}\\
&&\hspace*{-1.3cm}\frac{\partial n_i}{\partial t}+\frac{\partial}{\partial x}(n_i u_i)=0,
\label{1eq:3}\\
&&\hspace*{-1.3cm}\frac{\partial u_i}{\partial t}+ u_i\frac{\partial u_i}{\partial x}+\alpha_3n_i\frac{\partial n_i}{\partial x}=-\frac{\partial\phi}{\partial x},
\label{1eq:4}\\
&&\hspace*{-1.3cm}\frac{\partial^{2} \phi}{\partial x^{2}}=\alpha_4n_d+(1-\alpha_4)n_e-n_i,
\label{1eq:5}\
\end{eqnarray}
where $n_d$, $n_e$, and $n_i$ are normalized by $n_{d0}$, $n_{e0}$, and $n_{i0}$, respectively; the dust and ion fluid speed $u_d$ and $u_i$ are
normalized by DIAWs speed $C_i=(Z_ik_BT_e/m_i)^{1/2}$ (where $k_B$ is the Boltzmann constant);
the electrostatic wave potential $\phi$ is normalized by $k_BT_e/e$; the time and space are
normalized by the $\omega_{Pi}^{-1}=({m_i}/4\pi e^{2}Z_i^{2}n_{i0})^{1/2}$
and $\lambda_{Di} = (k_BT_e/4\pi e^{2}Z_in_{i0})^{1/2}$, respectively. The pressure term for the dust grains and ion
can be written as $P_d= P_{d0} (N_d/n_{d0})^{\gamma}$ and $P_i= P_{i0} (N_i/n_{i0})^{\gamma}$, respectively
[where $P_{d0}=n_{d0}k_B T_d$ ($P_{i0}=n_{i0}k_B T_i$) represents the equilibrium pressure associated with
the warm negative dust grains (positive ions), and $\gamma=(N + 2)/N$ and $\gamma=3$ (for one dimensional case, i.e., $N=1$)].
Other relevant physical parameters can be written as $\alpha_1=3m_iT_d/Z_im_dT_e$, $\alpha_2=\nu\mu$
(where $\nu=Z_d/Z_i$ and $\mu=m_i/m_d$), $\alpha_3=3T_i/Z_iT_e$, and $\alpha_4=Z_dn_{d0}/Z_in_{i0}$.
The normalized form of the number density of electron regarding the $\kappa$-distribution is represented as \cite{Chowdhury2019,Emamuddin2018}
\begin{eqnarray}
&&\hspace*{-1.3cm}n_e=\left[1 -\frac{e \phi}{(\kappa-3/2)}\right]^{-\kappa+\frac{1}{2}}.
\label{1eq:6}\
\end{eqnarray}
The parameter $\kappa$ denotes the super-thermality of electrons. Now, replacing Eq. \eqref{1eq:6}
into Eq. \eqref{1eq:5} and expanding up to third order in $\phi$, we get
\begin{eqnarray}
&&\hspace*{-1.3cm}\frac{\partial^2\phi}{\partial x^2}+n_i=1-\alpha_4+\alpha_4n_d+T_1\phi+T_2\phi^2+T_3\phi^3+\cdot\cdot\cdot,
\label{1eq:7}\
\end{eqnarray}
where
\begin{eqnarray}
&&\hspace*{-1.3cm}T_1 = \frac{(1-\alpha_4)(2\kappa-1)}{(2\kappa-3)},
\nonumber\\
&&\hspace*{-1.3cm}T_2 = \frac{(1-\alpha_4)(2\kappa-1)(2\kappa+1)}{2(2\kappa-3)^2},
\nonumber\\
&&\hspace*{-1.3cm}T_3 = \frac{(1-\alpha_4)(2\kappa-1)(2\kappa+1)(2\kappa+3)}{6(2\kappa-3)^3}.
\nonumber\
\end{eqnarray}
It should be noted here that the terms containing $T_1$, $T_2$, and $T_3$ at the
right hand side of Eq. \eqref{1eq:7} are due to the contribution of
super-thermal electrons.
\section{Derivation of the NLSE}
\label{1sec:Derivation of the NLSE}
To study  the MI of the DIAWs, first we want to derive the NLSE by employing the reductive perturbation method.
In that case, the stretched co-ordinates can be written in the following form \cite{Chowdhury2018}:
\begin{eqnarray}
&&\hspace*{-1.3cm}\xi={\epsilon}(x-v_g t),
\label{1eq:8}\\
&&\hspace*{-1.3cm}\tau={\epsilon}^2 t,
\label{1eq:9}\
\end{eqnarray}
where $v_g$ is the group speed and $\epsilon$ is a small parameter. After that the
dependent variables can be represented as \cite{Chowdhury2018}
\begin{eqnarray}
&&\hspace*{-1.3cm}\Upsilon(x,t)=\Upsilon_0+\sum_{m=1}^{\infty}\epsilon^{m}\sum_{l=-\infty}^{\infty}\Upsilon_{l}^{(m)}(\xi,\tau)\mbox{exp}[i l(kx-\omega t)].
\label{1eq:10}\
\end{eqnarray}
Here $\Upsilon$ = [$n_d$, $u_d$, $n_i$, $u_i$, $\phi$], $\Upsilon_0$ = [1, 0, 1, 0, 0]$^T$, and $\Upsilon_{l}^{(m)}$ = [$n_{dl}^{(m)}$, $u_{dl}^{(m)}$, $n_{il}^{(m)}$, $u_{il}^{(m)}$, $\phi_l^{(m)}$].
The carrier wave number (frequency) is defined as $k$ ($\omega$).
We can write the derivative operators as \cite{Chowdhury2018}
\begin{eqnarray}
&&\hspace*{-1.3cm}\frac{\partial}{\partial t}\rightarrow\frac{\partial}{\partial t}-\epsilon v_g\frac{\partial}{\partial \xi}+\epsilon^2\frac{\partial}{\partial\tau},
\label{1eq:11}\\
&&\hspace*{-1.3cm}\frac{\partial}{\partial x}\rightarrow\frac{\partial}{\partial x}+\epsilon\frac{\partial}{\partial \xi}.
\label{1eq:12}
\end{eqnarray}
Now, by substituting Eqs. \eqref{1eq:10}$-$\eqref{1eq:12} into Eqs. \eqref{1eq:1}$-$\eqref{1eq:4},
and \eqref{1eq:7}, and collecting the power terms of $\epsilon$, the first order ($m=1$ with $l=1$)
reduced equations can be written as
\begin{eqnarray}
&&\hspace*{-1.3cm}n_{d1}^{(1)}=\frac{\alpha_2k^2}{\alpha_1k^{2}-\omega^{2}}\phi_1^{(1)},
\label{1eq:13}\\
&&\hspace*{-1.3cm}u_{d1}^{(1)}=\frac{k\alpha_2\omega}{\alpha_1k^{2}-\omega^{2}}\phi_1^{(1)},
\label{1eq:14}\\
&&\hspace*{-1.3cm}n_{i1}^{(1)}=\frac{k^2}{\omega^{2}-\alpha_3k^{2}}\phi_1^{(1)},
\label{1eq:15}\\
&&\hspace*{-1.3cm}u_{i1}^{(1)}=\frac{k\omega}{\omega^{2}-\alpha_3k^{2}}\phi_1^{(1)}.
\label{1eq:16}\
\end{eqnarray}
These equations provide the dispersion relation of DIAWs in the following form
\begin{eqnarray}
&&\hspace*{-1.3cm}\omega^2\equiv\omega_f^2=\frac{k^{2}A+k^{2}\sqrt{A^2-4(k^{2}+T_1)B}}{2(k^{2}+T_1)},
\label{1eq:17}\\
&&\hspace*{-1.3cm}\omega^2\equiv\omega_s^2=\frac{k^{2}A-k^{2}\sqrt{A^2-4(k^{2}+T_1)B}}{2(k^{2}+T_1)},
\label{1eq:18}\
\end{eqnarray}
\begin{figure}[t!]
\centering
\includegraphics[width=80mm]{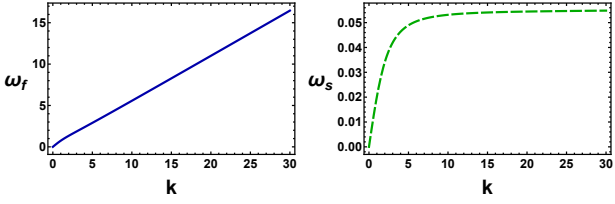}
\caption{Plot of $\omega_f$ vs $k$ (left panel), and  $\omega_s$ vs $k$ (right panel) when
$\kappa=1.7$, $\alpha_1=3\times10^{-8}$, $\alpha_3=0.3$, $\alpha_4=0.5$, $\nu=2\times 10^{3}$, and $\mu=3\times 10^{-6}$.}
\label{1Fig:F1}\
\end{figure}
where $\omega_f$ ($\omega_s$) is the fast (slow) mode of the DIAWs, $A=\alpha_1k^{2}+\alpha_3k^{2}+\alpha_1T_1+\alpha_3T_1+\alpha_2\alpha_4+1$,  and $B=\alpha_1+\alpha_1\alpha_3T_1+\alpha_1\alpha_3k^{2}+\alpha_2\alpha_3\alpha_4$.
We have  graphically observed the variation of the fast mode (left panel) and slow mode (right panel) with $k$ in Fig. \ref{1Fig:F1}.
It is clear from this figure that (a) the angular frequency of the DIAWs increases with carrier wave number (left panel);
(b) but $\omega_s$ exponentially increases with $k$ up to a particular value of $k$, and then it becomes
saturated showing no change  with the variation of
$k$ (right panel). Now, the second-order ($m=2$ with $l=1$) equations can be written as
\begin{eqnarray}
&&\hspace*{-1.3cm}n_{d1}^{(2)}=\frac{\alpha_2k^{2}}{\alpha_1k^{2}-\omega^{2}}\phi_1^{(2)}+\frac{i\mathcal I_1}{({\alpha_1k^{2}-\omega^{2})}^{2}}\frac{\partial \phi_1^{(1)}}{\partial\xi},
\label{1eq:19}\\
&&\hspace*{-1.3cm}u_{d1}^{(2)}=\frac{\alpha_2k\omega}{\alpha_1k^{2}-\omega^{2}}\phi_1^{(2)}+\frac{i\mathcal I_2}{({\alpha_1k^{2}-\omega^{2})}^{2}}\frac{\partial \phi_1^{(1)}}{\partial\xi},
\label{1eq:20}\\
&&\hspace*{-1.3cm}n_{i1}^{(2)}=\frac{k^{2}}{\omega^{2}-\alpha_3k^{2}}\phi_1^{(2)}-\frac{i\mathcal I_3}{({\omega^{2}-\alpha_3k^{2}})^{2}}\frac{\partial \phi_1^{(1)}}{\partial\xi},
\label{1eq:21}\\
&&\hspace*{-1.3cm}u_{i1}^{(2)}=\frac{k\omega}{\omega^{2}-\alpha_3k^{2}}\phi_1^{(2)}-\frac{i\mathcal I_4}{({\omega^{2}-\alpha_3k^{2})}^{2}}\frac{\partial \phi_1^{(1)}}{\partial\xi},
\label{1eq:22}\
\end{eqnarray}
where
\begin{eqnarray}
&&\hspace*{-1.3cm}\mathcal I_1 =\alpha_2k\omega^{2}-2v_g\alpha_2k^{2}\omega-\alpha_2k(\alpha_1k^{2}-\omega^{2})+\alpha_1\alpha_2k^{3},
\nonumber\\
&&\hspace*{-1.3cm}\mathcal I_2 =\alpha_2\omega^{3}-2v_g\alpha_2k\omega^{2}+\alpha_1\alpha_2k^{2}\omega-v_g\alpha_2k(\alpha_1k^{2}-\omega^{2}),
\nonumber\\
&&\hspace*{-1.3cm}\mathcal I_3 =\alpha_3k^{3}-2v_gk^{2}\omega+k\omega^{2}+k(\omega^{2}-\alpha_3k^{2}),
\nonumber\\
&&\hspace*{-1.3cm}\mathcal I_4 =\alpha_3\omega k^{2}-2v_gk\omega^{2}+v_gk(\omega^{2}-\alpha_3k^{2})+\omega^{3}.
\nonumber\
\end{eqnarray}
The second-order ($m=2$ with $l=1$) equations
and with the compatibility condition, we can write the group velocity ($v_g$) of DIAWs in the following form
\begin{eqnarray}
&&\hspace*{-1.3cm}v_g=\frac{k^2\omega^4(\alpha_3+\alpha_1\alpha_2\alpha_4)+k^4\omega^2(\alpha_1^2+\alpha_2\alpha_3^2\alpha_4)+\mathcal I_5}{2k\omega[(\alpha_1k^{2}-\omega^{2})^{2}+\alpha_2\alpha_4{(\omega^{2}-\alpha_3k^{2})}^{2}]},
\label{1eq:23}\
\end{eqnarray}
where
\begin{eqnarray}
&&\hspace*{-1.2cm}\mathcal I_5 =\alpha_1\alpha_3k^6(\alpha_1+\alpha_2\alpha_3\alpha_4)-2\alpha_1\alpha_3k^4\omega^2(1+\alpha_2\alpha_4)
\nonumber\\
&&\hspace*{-0.5cm}-2k^2\omega^4(\alpha_1+\alpha_2\alpha_3\alpha_4)+\omega^6(1+\alpha_2\alpha_4)
\nonumber\\
&&\hspace*{-0.5cm}+(\alpha_1k^2-\omega^2)(\omega^2-\alpha_3k^2)[k^2(\alpha_1+\alpha_2\alpha_3\alpha_4)
\nonumber\\
&&\hspace*{-0.5cm}-\omega^2(1+\alpha_2\alpha_4)-2(\alpha_1k^2-\omega^2)(\omega^2-\alpha_3k^2)].
\nonumber\
\end{eqnarray}
The coefficients of $\epsilon$ for $m=2$ and $l=2$ provide the second order harmonic amplitudes
which are found to be proportional to $|\phi_1^{(1)}|^2$
\begin{eqnarray}
&&\hspace*{-1.3cm}n_{d2}^{(2)}=T_4|\phi_1^{(1)}|^2,
\label{1eq:24}\\
&&\hspace*{-1.3cm}u_{d2}^{(2)}=T_5 |\phi_1^{(1)}|^2,
\label{1eq:25}\\
&&\hspace*{-1.3cm}n_{i2}^{(2)}=T_6|\phi_1^{(1)}|^2,
\label{1eq:26}\\
&&\hspace*{-1.3cm}u_{i2}^{(2)}=T_7 |\phi_1^{(1)}|^2,
\label{1eq:27}\\
&&\hspace*{-1.3cm}\phi_{2}^{(2)}=T_8 |\phi_1^{(1)}|^2,
\label{1eq:28}\
\end{eqnarray}
where
\begin{eqnarray}
&&\hspace*{-1.3cm}T_4=\frac{\alpha_1\alpha_2k^6(2\alpha_1T_8-\alpha_2)-\alpha_2k^4\omega^2(4\alpha_1T_8+3\alpha_2)-\alpha_1\alpha_2^2k^6}{2(\alpha_1k^{2}-\omega^{2})^{3}},
\nonumber\\
&&\hspace*{-1.3cm}T_5=\frac{k^4\omega(T_4\alpha_1^2-\alpha_2^2)-2\alpha_1T_4k^2\omega^3+T_4\omega^5}{k(\alpha_1k^{2}-\omega^{2})^{2}},
\nonumber\\
&&\hspace*{-1.3cm}T_6=\frac{\alpha_3k^6(2\alpha_3T_8+1)-k^4\omega^2(4\alpha_3T_8-3)+2T_8k^2\omega^4}{2(\omega^{2}-\alpha_3k^{2})^{3}},
\nonumber\\
&&\hspace*{-1.3cm}T_7=\frac{k^4\omega(T_6\alpha_3^2-1)-2\alpha_3T_6k^2\omega^3+T_6\omega^5}{k(\omega^{2}-\alpha_3k^{2})^{2}},
\nonumber\\
&&\hspace*{-1.3cm}T_8=\frac{2T_2(\alpha_1k^{2}-\omega^{2})^{3}(\omega^{2}-\alpha_3k^{2})^{3}-\mathcal O_1}{ 2k^{2}(\omega^{2}-\alpha_3k^{2})^{2}(\alpha_1k^{2}-\omega^{2})^{3}-\mathcal O_2},
\nonumber\
\end{eqnarray}
where
\begin{eqnarray}
&&\hspace*{-1.3cm}\mathcal O_1=\alpha_4\alpha_2^{2}k^{4}(\omega^{2}-\alpha_3k^{2})^{3}(\alpha_1k^{2}+3\omega^{2})
\nonumber\\
&&\hspace*{-0.5cm}+(\alpha_1k^{2}-\omega^{2})^{3}k^{4}(\alpha_3k^{2}+3\omega^{2}),
\nonumber\\
&&\hspace*{-1.3cm}\mathcal O_2 =2(\alpha_1k^{2}-\omega^{2})^{3}(\omega^{2}-\alpha_3k^{2})^{3}(4k^{2}+T_1)
\nonumber\\
&&\hspace*{-0.5cm}+2\alpha_2\alpha_4k^{2}(\alpha_1k^{2}-\omega^{2})^{2}(\omega^{2}-\alpha_3k^{2})^{3}.
\nonumber\
\end{eqnarray}
Now, we consider the expression for ($m=3$ with $l=0$) and ($m=2$ with $l=0$),
which leads the zeroth harmonic modes. Thus, we obtain
\begin{eqnarray}
&&\hspace*{-1.3cm}n_{d0}^{(2)}=T_{9}|\phi_1^{(1)}|^2,
\label{1eq:29}\\
&&\hspace*{-1.3cm}u_{d0}^{(2)}=T_{10}|\phi_1^{(1)}|^2,
\label{1eq:30}\\
&&\hspace*{-1.3cm}n_{i0}^{(2)}=T_{11}|\phi_1^{(1)}|^2,
\label{1eq:31}\\
&&\hspace*{-1.3cm}u_{i0}^{(2)}=T_{12}|\phi_1^{(1)}|^2,
\label{1eq:32}\\
&&\hspace*{-1.3cm}\phi_0^{(2)}=T_{13}|\phi_1^{(1)}|^2,
\label{1eq:33}\
\end{eqnarray}
where
\begin{eqnarray}
&&\hspace*{-1.3cm}T_9=\frac{\alpha_1\alpha_2k^4(\alpha_2-T_{13}\alpha_1)+O_3-\alpha_2T_{13}\omega^4}
{(v_g^{2}-\alpha_1)(\alpha_1k^{2}-\omega^{2})^{2}},
\nonumber\\
&&\hspace*{-1.3cm}T_{10}=\frac{v_gT_9\alpha_1^2k^4-2\alpha_2^2k^3\omega-2v_gT_9\alpha_1k^2\omega^2+v_gT_9\omega^4}{(\alpha_1k^{2}-\omega^{2})^{2}},
\nonumber\\
&&\hspace*{-1.3cm}T_{11}=\frac{\alpha_3k^4(T_{13}\alpha_3+1)+2v_gk^3\omega-k^2\omega^2(2T_{13}\alpha_3-1)+T_{13}\omega^4}{(v_g^{2}-\alpha_3)(\omega^{2}-\alpha_3k^{2})^{2}},
\nonumber\\
&&\hspace*{-1.3cm}T_{12}=\frac{v_gT_{11}\alpha_3^2k^4-2k^3\omega-2v_gT_{11}\alpha_3k^2\omega^2+v_gT_{11}\omega^4}{(\omega^{2}-\alpha_3k^{2})^{2}},
\nonumber\\
&&\hspace*{-1.3cm}T_{13}=\frac{2T_2(v_g^{2}-\alpha_1)(v_g^{2}-\alpha_3)(\alpha_1k^{2}-\omega^{2})^{2}(\omega^{2}-\alpha_3k^{2})^{2}+\mathcal O_4}{(\alpha_1k^{2}-\omega^{2})^{2}(\omega^{2}-\alpha_3k^{2})^{2}\times\mathcal O_5},
\nonumber\
\end{eqnarray}
where
\begin{eqnarray}
&&\hspace*{-1.3cm}\mathcal O_3=2v_g\alpha_2^2k^3\omega+\alpha_2k^2\omega^2(\alpha_2+2T_{13}\alpha_1),
\nonumber\\
&&\hspace*{-1.3cm}\mathcal O_4=(v_g^{2}-\alpha_3)(\omega^{2}-\alpha_3k^{2})^{2}\alpha_4\alpha_2^{2}k^{2}(2v_gk\omega+\alpha_1k^{2}+\omega^{2})
\nonumber\\
&&\hspace*{-0.5cm}-k^{2}(\alpha_1k^{2}-\omega^{2})^{2}(v_g^{2}-\alpha_1)(2v_gk\omega+\alpha_3k^{2}+\omega^{2}),
\nonumber\\
&&\hspace*{-1.3cm}\mathcal O_5=(v_g^{2}-\alpha_1)-T_1(v_g^{2}-\alpha_1)(v_g^{2}-\alpha_3)+\alpha_2\alpha_4(v_g^{2}-\alpha_3).
\nonumber\
\end{eqnarray}
Finally, the third harmonic modes ($m=3$) and ($l=1$), with the help of
Eqs. \eqref{1eq:13}-\eqref{1eq:33}, give a set of equations, which can be
reduced to the following NLSE:
\begin{eqnarray}
&&\hspace*{-1.3cm}i\frac{\partial\Phi}{\partial\tau}+P\frac{\partial^2\Phi}{\partial\xi^2}+Q|\Phi|^2\Phi=0,
\label{1eq:34}\
\end{eqnarray}
where $\Phi=\phi_1^{(1)}$ has been taken for simplicity. In Eq. \eqref{1eq:34}, $P$ is the dispersion coefficient which can be written as
\begin{eqnarray}
&&\hspace*{-1.3cm}P=\frac{\mathcal O_6}{2\omega(\alpha_1k^{2}-\omega^{2})(\omega^{2}-\alpha_3k^{2})k^{2}\times\mathcal O_7},
\nonumber\
\end{eqnarray}
where
\begin{eqnarray}
&&\hspace*{-1.3cm}\mathcal O_6=(\alpha_1k^{2}-\omega^{2})^{3}[(v_gk\omega-\alpha_3k^{2})\{2v_gk\omega-\alpha_3k^{2}-\omega^{2}
\nonumber\\
&&\hspace*{-0.5cm}-(\omega^{2}-\alpha_3k^{2})\}+(v_gk-\omega)\{2v_gk\omega^{2}-v_gk(\omega^{2}-\alpha_3k^{2})
\nonumber\\
&&\hspace*{-0.5cm}-\alpha_3k^{2}\omega-\omega^{3}\}]-\alpha_2\alpha_4(\omega^{2}-\alpha_3k^{2})^{3}[(v_gk\omega-\alpha_1k^{2})
\nonumber\\
&&\hspace*{-0.5cm}\{2v_gk\omega+(\alpha_1k^{2}-\omega^{2})-\alpha_1k^{2}-\omega^{2}\}+(v_gk-\omega)
\nonumber\\
&&\hspace*{-0.5cm}\{2v_gk\omega ^{2}-\alpha_1\omega k^{2}-\omega^{3}+v_gk(\alpha_1k^{2}-\omega^{2})\}
\nonumber\\
&&\hspace*{-0.5cm}-(\alpha_1k^{2}-\omega^{2})^{3}(\omega^{2}-\alpha_3k^{2})^{3}],
\nonumber\\
&&\hspace*{-1.3cm}\mathcal O_7=(\alpha_1k^{2}-\omega^{2})^{2}+\alpha_2\alpha_4(\omega^{2}-\alpha_3k^{2})^{2},
\nonumber\
\end{eqnarray}
and $Q$ is the nonlinear coefficient which can be written as
\begin{eqnarray}
&&\hspace*{-1.3cm}Q=\frac{2T_2(T_8+T_{13})(\alpha_1k^{2}-\omega^{2})^{2}(\omega^{2}-\alpha_3k^{2})^{2}+\mathcal O_8}{2\omega k^{2}\{(\alpha_1k^{2}-\omega^{2})^{2}+\alpha_2\alpha_4(\omega^{2}-\alpha_3k^{2})^{2}\}},
\nonumber\
\end{eqnarray}
where
\begin{eqnarray}
&&\hspace*{-1.3cm}\mathcal O_8=3T_3(\alpha_1k^{2}-\omega^{2})^{2}(\omega^{2}-\alpha_3k^{2})^{2}-2k^{3}\omega(\alpha_1k^{2}-\omega^{2})^{2}
\nonumber\\
&&\hspace*{-0.5cm}(T_7+T_{12})-2\alpha_2\alpha_4k^{3}\omega(\omega^{2}-\alpha_3k^{2})^{2}(T_5+T_{10})
\nonumber\\
&&\hspace*{-0.5cm}-(\alpha_1k^{2}-\omega^{2})^{2}(k^{2}\omega^{2} +\alpha_3k^{4})(T_6+T_{11})
\nonumber\\
&&\hspace*{-0.5cm}-(\omega^{2}-\alpha_3k^{2})^{2}(\alpha_1\alpha_2\alpha_4k^{4}+\alpha_2\alpha_4k^{2}\omega^{2})(T_4+T_9).
\nonumber\
\end{eqnarray}
The space and time evolution of the DIAWs in a dusty plasma are directly governed by the
coefficients $P$ and $Q$, and indirectly governed by different plasma parameters
such as $\alpha_1$, $\alpha_3$, $\alpha_4$, $\mu$, $\nu$, $\kappa$, and $k$, etc.
 Thus, these plasma parameters can significantly modify the stability conditions
of DIAWs in a dusty plasma.
\begin{figure}[h]
\centering
\includegraphics[width=70mm]{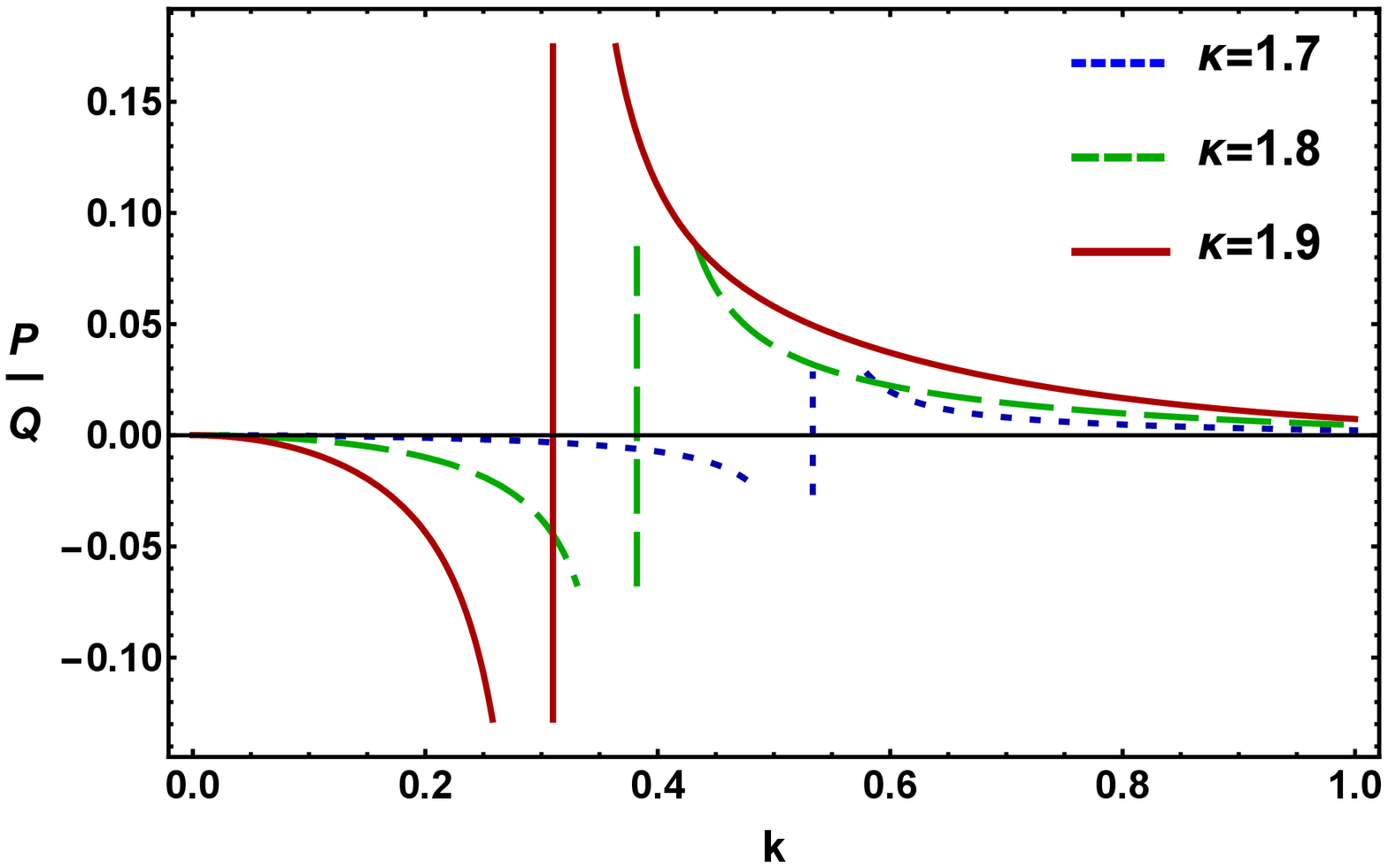}
\caption{Plot of $P/Q$ vs $k$ for different values of  $\kappa$ when $\alpha_1=3\times10^{-8}$, $\alpha_3=0.3$,
$\alpha_4=0.5$, $\nu=2\times 10^3$, $\mu=3\times 10^{-6}$, and $\omega\equiv\omega_f$.}
\label{1Fig:F2}\
\vspace{0.8cm}
\includegraphics[width=70mm]{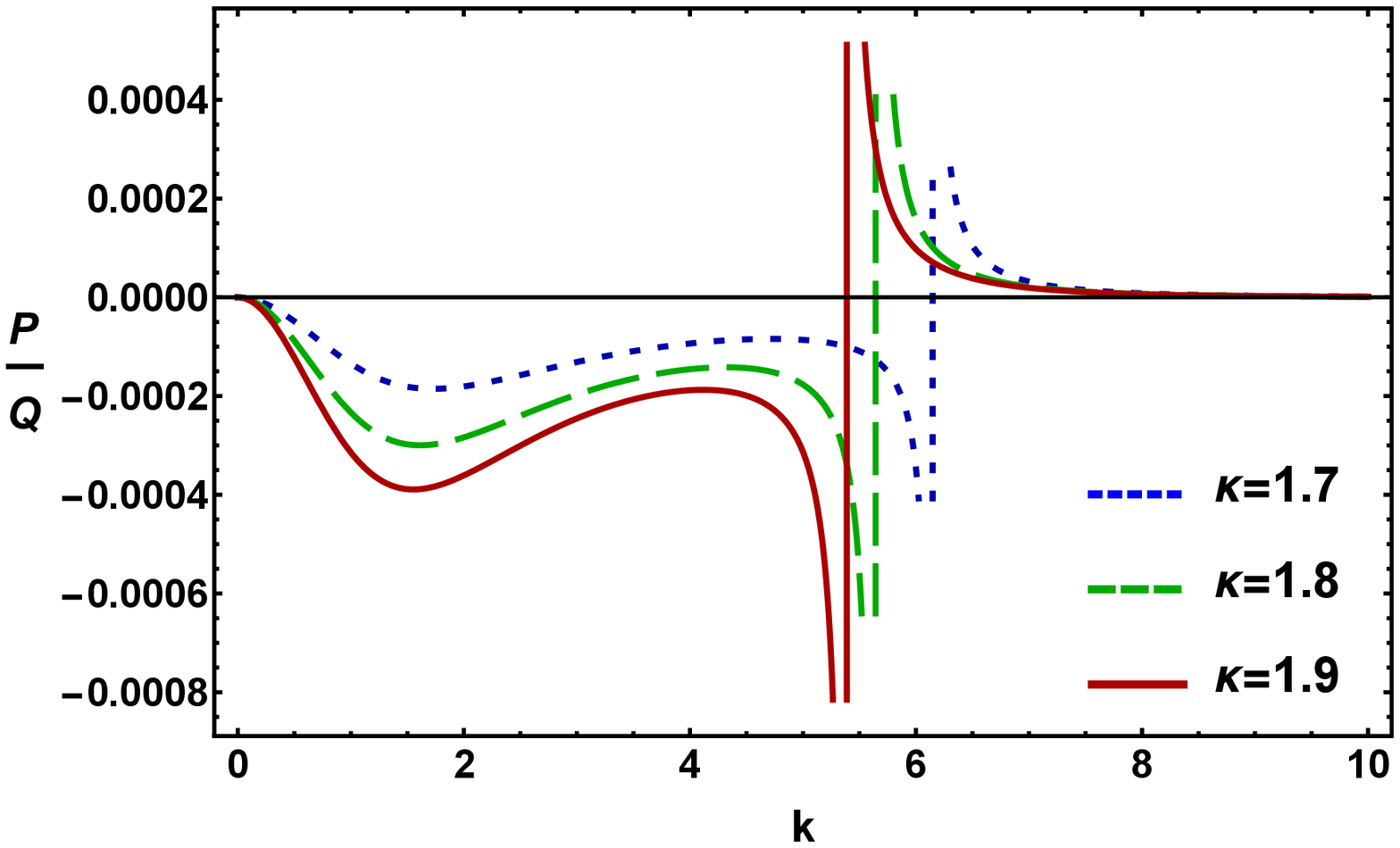}
\caption{Plot of $P/Q$ vs $k$ for different values of  $\kappa$
when $\alpha_1=3\times10^{-8}$, $\alpha_3=0.3$, $\alpha_4=0.5$, $\nu=2\times 10^{3}$,
 $\mu=3\times 10^{-6}$, and $\omega\equiv\omega_s$.}
\label{1Fig:F3}\
\end{figure}
\begin{figure}[h]
\centering
\includegraphics[width=70mm]{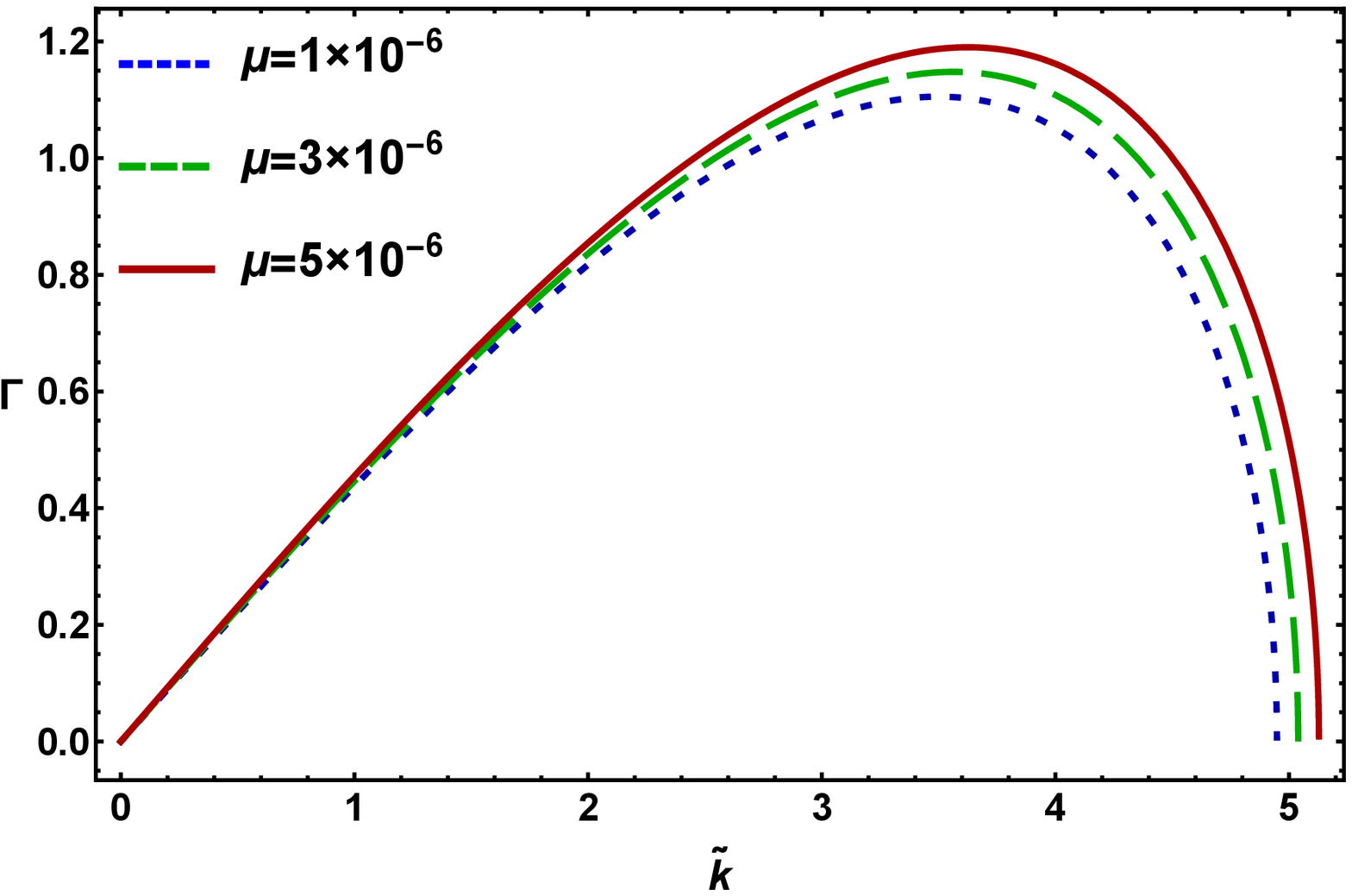}
\caption{Plot of $\Gamma$ vs $\tilde{k}$ for different values of $\mu$ when $\kappa=1.7$,
$\alpha_1=3\times10^{-8}$, $\alpha_3=0.3$, $\alpha_4=0.5$, $\nu=2\times 10^{3}$, and $\omega\equiv\omega_f$.}
\label{1Fig:F4}\
\vspace{0.8cm}
\includegraphics[width=70mm]{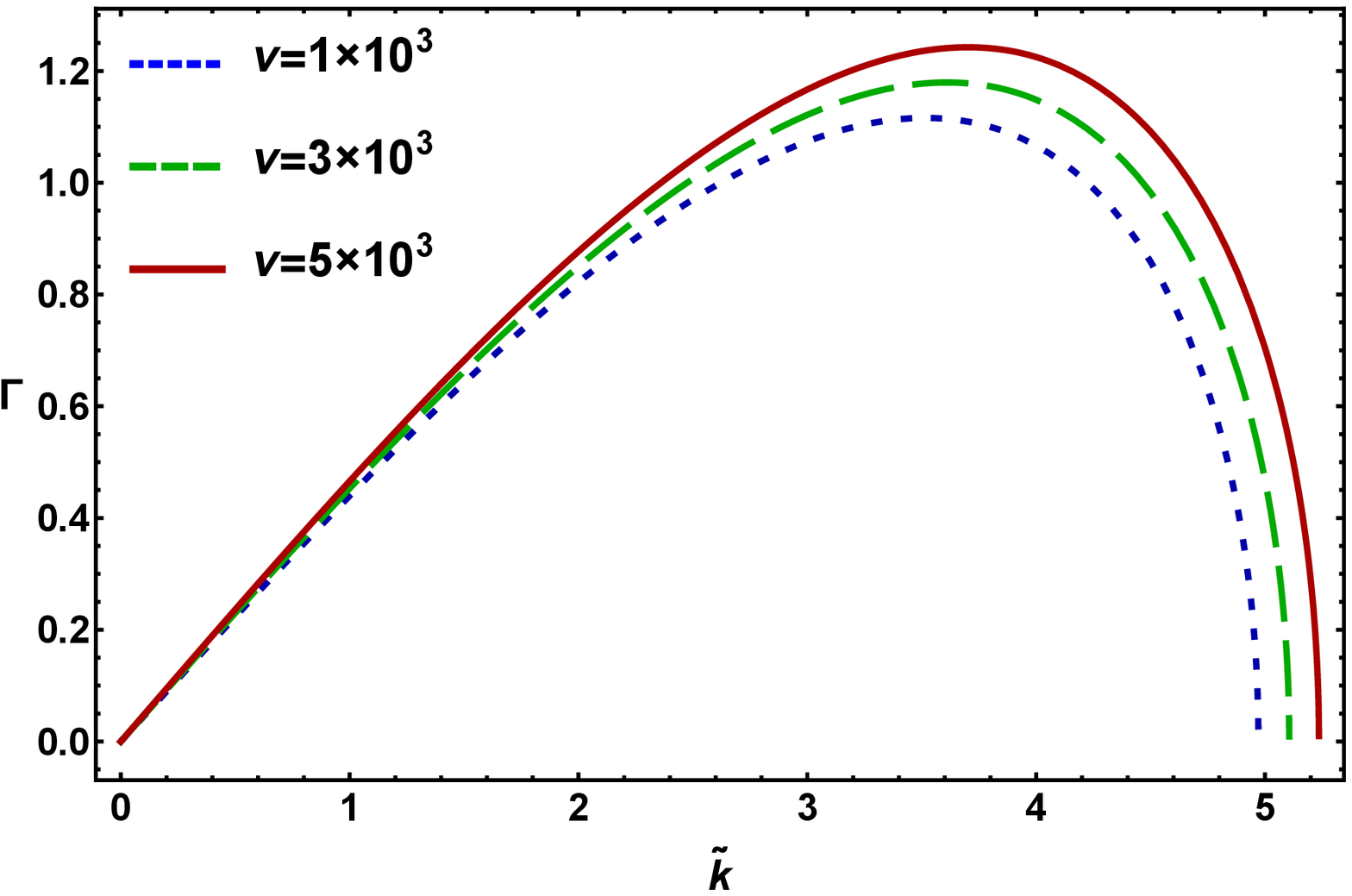}
\caption{Plot of $\Gamma$ vs $\tilde{k}$ for different values of $\nu$, when $\kappa=1.7$, $\alpha_1=3\times10^{-8}$, $\alpha_3=0.3$, $\alpha_4=0.5$, $\mu=3\times 10^{-6}$, and $\omega\equiv\omega_f$.}
\label{1Fig:F5}\
\end{figure}
\begin{figure}[h]
\centering
\includegraphics[width=70mm]{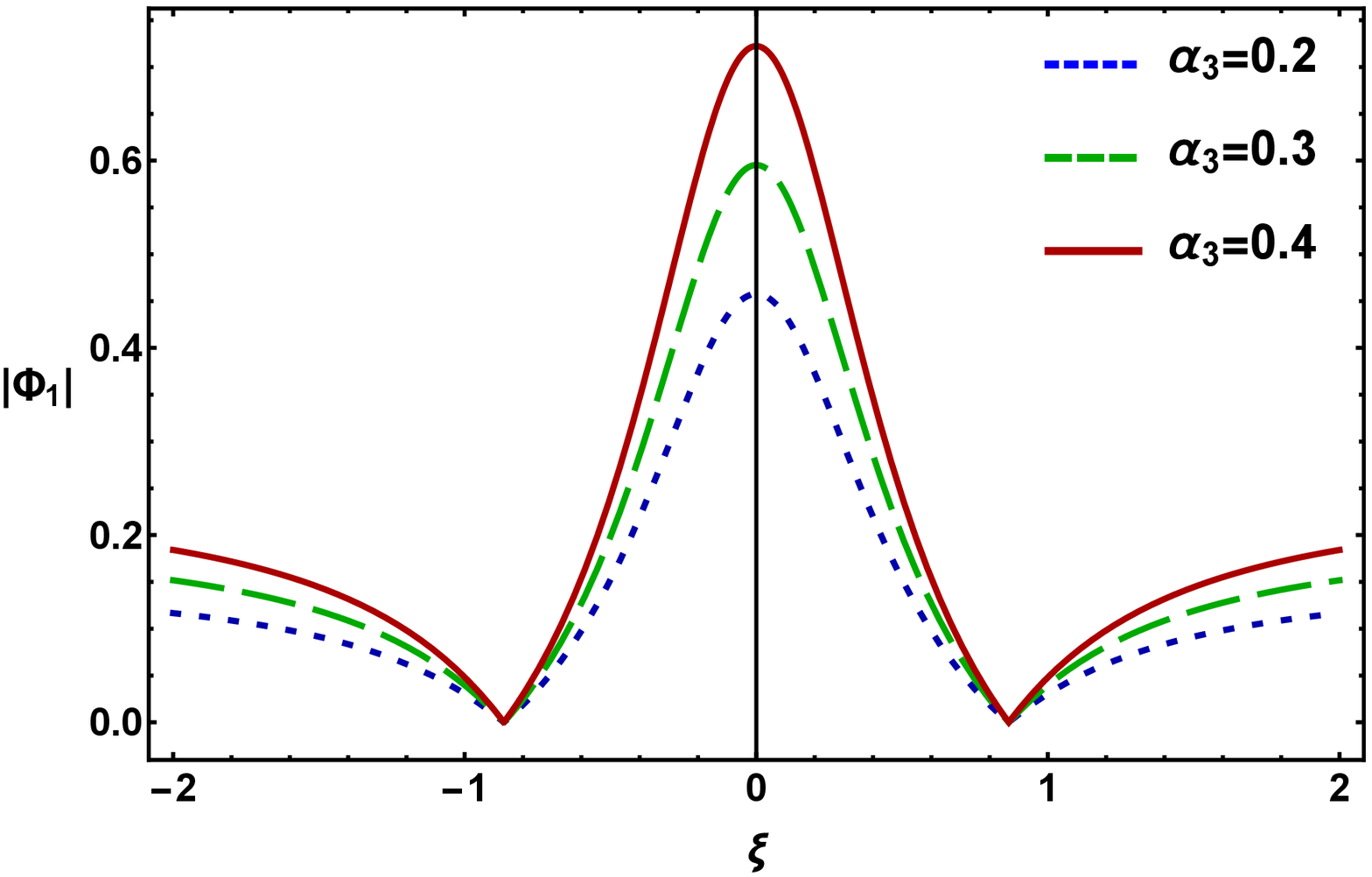}
\caption{Plot of $|\Phi_1|$ vs $\xi$ for different values of $\alpha_3$,
when $\kappa=1.7$, $\alpha_1=3\times10^{-8}$, $\alpha_4=0.5$, $\nu=2\times 10^3$, $\mu=3\times 10^{-6}$, and $\omega\equiv\omega_f$.}
\label{1Fig:F6}\
\end{figure}
\begin{figure}[h]
\centering
\includegraphics[width=70mm]{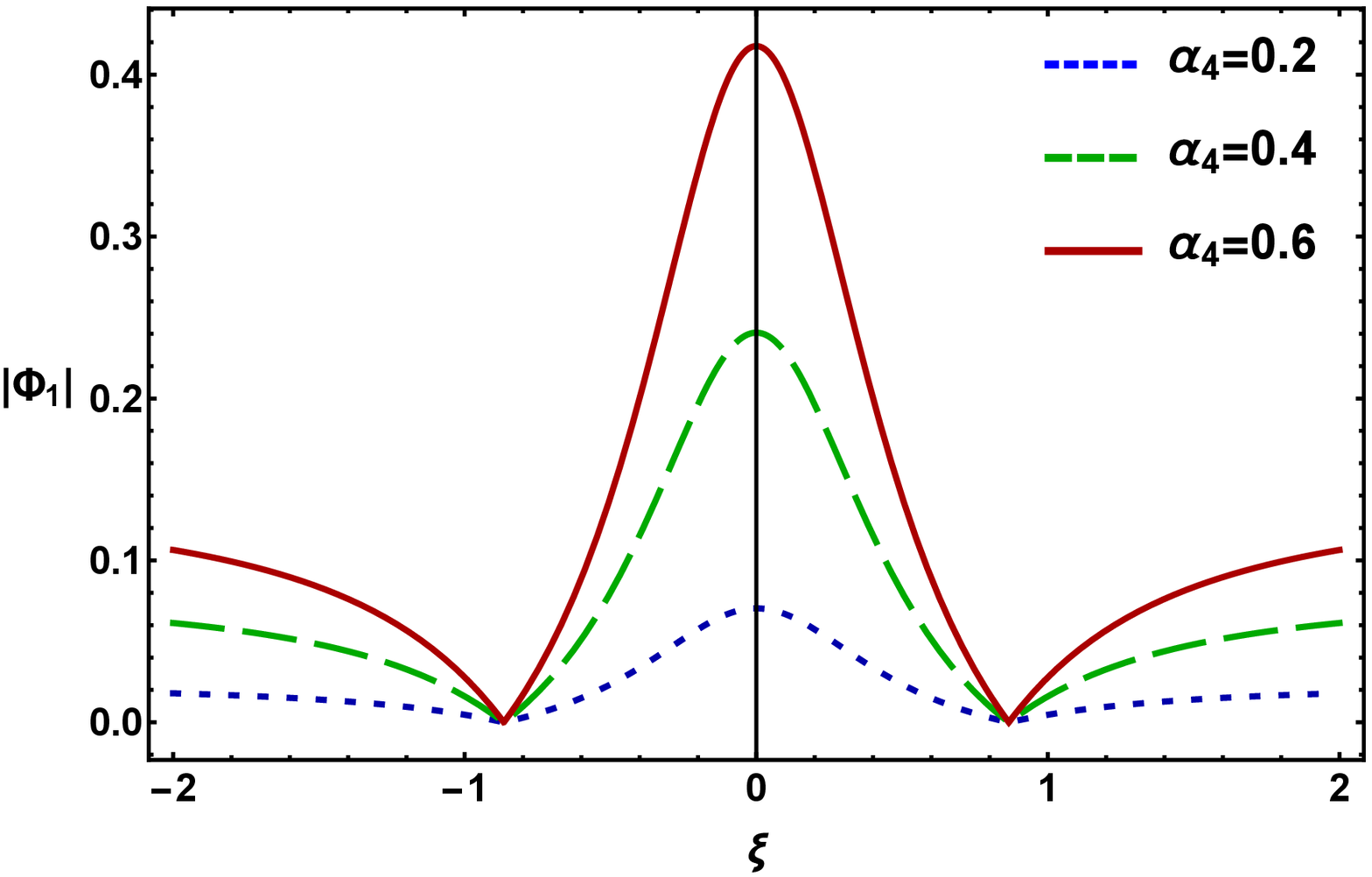}
\caption{Plot of $|\Phi_1|$ vs $\xi$ for different values of $\alpha_4$,
when $\kappa=1.7$, $\alpha_1=3\times10^{-8}$, $\alpha_3=0.3$, $\nu=2\times 10^3$, $\mu=3\times 10^{-6}$, and $\omega\equiv\omega_f$.}
\label{1Fig:F7}\
\end{figure}
\begin{figure}[h]
\centering
\includegraphics[width=70mm]{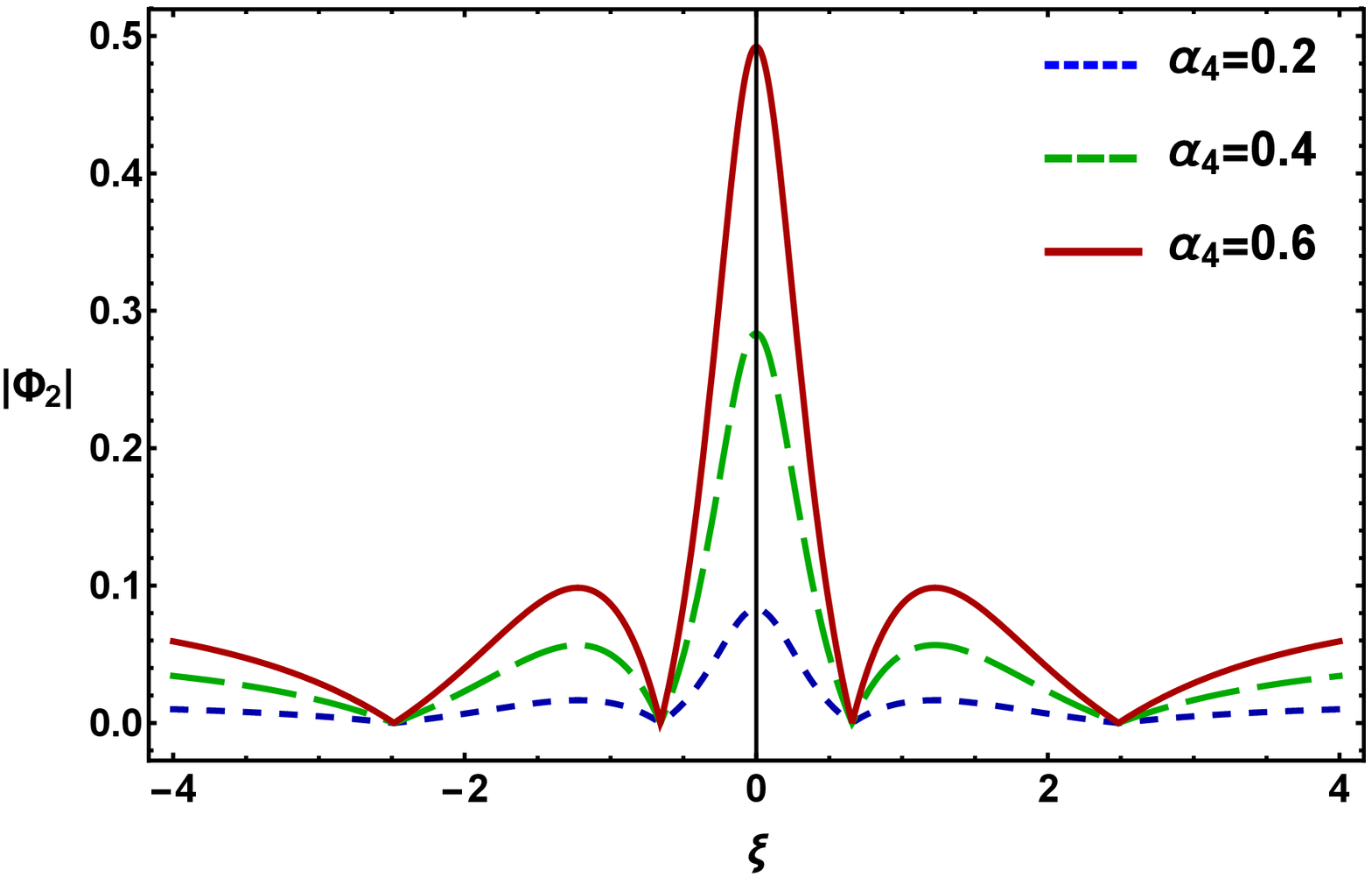}
\caption{Plot of $|\Phi_2|$ vs $\xi$ for different values of $\alpha_4$, when $\kappa=1.7$, $\alpha_1=3\times10^{-8}$, $\alpha_3=0.3$, $\nu=2\times 10^3$, $\mu=3\times 10^{-6}$, and $\omega\equiv\omega_f$.}
\label{1Fig:F8}\
\end{figure}
\section{Modulational instability and rogue waves}
\label{1sec:Modulational instability and rogue waves}
The stable and unstable parametric regimes of the DIAWs have been determined by the sign of the dispersion ($P$)
and nonlinear ($Q$) coefficients of the standard NLSE \cite{Chowdhury2018,Kourakis2005,Fedele2002,Sultana2011}.
When $P$ and $Q$ have same sign (i.e., $P/Q>0$),
the evolution of the DIAWs amplitude is modulationally unstable. On the other hand, when $P$ and $Q$ have
opposite sign (i.e., $P/Q<0$), the DIAWs are modulationally stable in the presence of external perturbations.
The plot of $P/Q$ against $k$ yields stable and unstable parametric regimes of DIAWs.
The point, at which transition of $P/Q$ curve intersects with $k$-axis, is known as threshold
or critical wave number $k$ ($=k_c$) \cite{Chowdhury2018,Kourakis2005,Fedele2002,Sultana2011}.
When $P/Q>0$ and $\tilde{k}<k_c$, the MI growth rate $(\Gamma)$ is given by \cite{Chowdhury2018,Sultana2011}
\begin{eqnarray}
&&\hspace*{-1.3cm}\Gamma=|P|~{\tilde{k}^{2}}\sqrt{\frac{k^{2}_{c}}{\tilde{k}^{2}}-1},
\label{1eq:35}
\end{eqnarray}
where $\tilde{k}$ is the modulated wave number. The first-order rational solution of NLSE in
the unstable parametric regime (i.e., $P/Q>0$) is given by \cite{Akhmediev2009,Akhmediev2009a,Guo2013}
\begin{eqnarray}
&&\hspace*{-1.3cm}\Phi_1 (\xi, \tau)=\sqrt{\frac{2P}{Q}}\Big[\frac{4+16 i\tau P}{1+4 \xi^{2} + 16\tau^{2} P^{2}}-1\Big] \mbox{exp} (2i\tau P).
\label{1eq:36}
\end{eqnarray}
The second-order RWs, which are the super-position of two or more first-order RWs, solution can be
written as \cite{Akhmediev2009,Akhmediev2009a,Guo2013}
\begin{eqnarray}
&&\hspace*{-1.3cm}\Phi_2(\xi,\tau)=\sqrt{\frac{P}{Q}}\Big[1+\frac{G_2(\xi,\tau)+iM_2(\xi,\tau)}{D_2(\xi,\tau)}\Big] \mbox{exp} (i\tau P),
\label{1eq:37}\
\end{eqnarray}
where $G_2$, $M_2$, and $D_2$ are some polynomials associated with the variables $\xi$ and $\tau$, which can be written as
\begin{eqnarray}
&&\hspace*{-1.3cm}G_2(\xi,\tau)=\frac{-\xi^{4}}{2}-6(P\xi\tau)^{2}-10(P\tau)^{4}-\frac{3\xi^{2}}{2}-9(P\tau)^{2}+\frac{3}{8},
\nonumber\\
&&\hspace*{-1.3cm}M_2(\xi,\tau)=-P\tau\Big[\xi^{4}+4(P\xi\tau)^{2}+4(P\tau)^{4}-3\xi^{2}+2(P\tau)^{2}-\frac{15}{4}\Big],
\nonumber\\
&&\hspace*{-1.3cm}D_2(\xi,\tau)=\frac{\xi^{6}}{12}+\frac{\xi^{4}(P\tau)^{2}}{2}+\xi^{2}(P\tau)^{4}
+\frac{\xi^{4}}{8}+\frac{9(P\tau)^{4}}{2}
\nonumber\\
&&\hspace*{0.3cm}-\frac{3(P\xi\tau)^{2}}{2}+\frac{9\xi^{2}}{16}+\frac{33(P\tau)^{2}}{8}+\frac{3}{32}.
\nonumber\
\end{eqnarray}
Eq. \eqref{1eq:36} and \eqref{1eq:37} represent the solution of the first and second-order DIARWs associated with DIAWs, respectively.
\section{Results and discussion}
\label{1sec:Results and discussion}
Generally, in dust-ion-acoustic waves, the positive
ion mass provides the moment of inertia and the thermal pressure of the electrons provides the
restoring force in the presence of immobile negative dust grains. On the other hand, in dust-acoustic waves,
the moment of inertia is provided by the massive negative dust grains and the restoring force is provided
by the thermal pressure of electrons and positive ions. But the consideration of thermal effects of the positive
ions can contribute substantively to the moment of inertia along with negative dust grains in the formation of the DIAWs.
In our present plasma model, we consider thermal effects of the ions along with negative dust grains (both ion and dust grains are
inertial) and would like to examine the contribution of inertial ions in the formation of the DIAWs.
We have also considered that $m_d=10^6m_i$, $Z_d=10^3Z_i$, and $T_e>T_i>T_d$.

We have graphically observed the variation of the $P/Q$ with respect to $k$ for different
values of $\kappa$ in Figs. \ref{1Fig:F2} and \ref{1Fig:F3} corresponding to the DIA fast ($\omega_f$) and
slow ($\omega_s$) modes when other plasma parameters are $\alpha_1=3\times10^{-8}$, $\alpha_3=0.3$,
$\alpha_4=0.5$, $\nu=2\times 10^{3}$, and $\mu=3\times 10^{-6}$. It has been observed that both modulationally
stable (i.e., $P/Q<0$) and unstable (i.e., $P/Q>0$) parametric regimes of the DIAWs can exist for both fast and slow
modes. The critical wave number ($k_c$), which divides the stable and parametric regimes, decreases with the increase of
$\kappa$ under the consideration of DIA fast and slow modes.

The effects of mass and charge state of the positive ion and negative dust grain on the MI of DIAWs in the
modulationally unstable parametric regime can be seen in Figs. \ref{1Fig:F4} and \ref{1Fig:F5}.
Figure \ref{1Fig:F4} describes the modification of the MI growth rate due to existence of heavy negative dust
grains and light positive ions. It is obvious from this figure that the $\Gamma$, initially, increases with $\tilde{k}$,
and becomes maximum for a particular value of $\tilde{k}$, then decreases to zero. The maximum value of the
growth rate increases with the mass of the positive ion but decreases with mass of the negative dust grain. On the
other hand , it is clear from Fig. \ref{1Fig:F5} that the maximum value of the MI growth rate increases (decreases)
with the increase in the value of the negative dust (positive ion) charge state. Physically, the nonlinearity as
well as the maximum value of the growth rate increases (decreases) with the increasing charge state of the negative dust grains (positive ions).
So, the mass and charge state of the positive ions and negative dust grains have to play an opposite role in the
dynamics of the plasma medium.

In our present investigation, we have considered the thermal effect of ions as well as
the moment of inertia of positive ions along with negative dust grains. So, it is important to
examine how the dynamics of the plasma system changes due the consideration of thermal effect as well
as the moment of inertia of positive ions along with negative dust grains. We have numerically analyzed
Eq. \eqref{1eq:36} in Fig. \ref{1Fig:F6} to observe the effects of ion temperature (via $\alpha_3$) in the formation of first-order DIARWs.
The amplitude and width of the first-order DIARWs increase with the increase in the value of the ion temperature ($T_i$) for
a fixed value of ion charge state ($Z_i$) and electron temperature ($T_e$). The physics behind this result is that
the nonlinearity as well as the amplitude of the first-order DIARWs increases with ion temperature. So, the consideration of the
thermal effect of the inertial positive ion significantly changes the dynamics of the plasma medium.
Figure \ref{1Fig:F7} illustrates the variation of the first-order DIARWs with different values of $\alpha_4$,
and it is clear from this figure that the amplitude and width of the first-order DIARWs increases with $\alpha_4$,
and this means that the charge state of the positive ion ($Z_i$) minimizes the nonlinearity as well as
the amplitude of the first-order DIARWs while the charge state of the negative dust grain ($Z_d$) maximizes
the nonlinearity as well as the amplitude of the first-order DIARWs for a constant value of $n_{d0}$ and $n_{i0}$.

The second-order DIARWs has been depicted by using Eq. \eqref{1eq:37} in Fig. \ref{1Fig:F8}, and it
is clear from this figure that the negative dust population ($n_{d0}$) enhances the amplitude of the second-order DIARWs
while the positive ion population ($n_{i0}$) reduces the amplitude of the second-order DIARWs
for their constant charge states (via $\alpha_4$). Similarly, the increasing charge state of the negative
dust grains (positive ions) enhances (reduces) the nonlinearity as well as amplitude of the second-order DIARWs when
their number density remains constant.
\section{Conclusion}
\label{1sec:Conclusion}
We have studied the stability conditions of the DIAWs in a three-component dusty plasma by considering
the thermal effect of the positive ions. In our present analysis the moment of inertia for the formation of the DIAWs is provided by
the positively charged warm ions and negatively charged dust grains, and the restoring force is provided by the
thermal pressure of the electrons. The evolution of the DIARWs associated with DIAWs is governed by the standard NLSE, and it is interesting
that the nonlinear and dispersive coefficients of the NLSE can easily predict the modulationally stable and unstable
parametric regimes of DIAWs. The results that have been found from our investigation can be summarized as follows:
\begin{itemize}
  \item Both modulationally stable (i.e., $P/Q<0$) and unstable (i.e., $P/Q>0$) parametric regimes of the DIAWs can exist for both fast and slow modes.
  \item The nonlinearity as well as the amplitude of the first-order DIARWs increases with ion temperature.
  \item The negative dust population ($n_{d0}$) enhances the amplitude of the second-order DIARWs
while the positive ion population ($n_{i0}$) reduces the amplitude of the second-order DIARWs
for their constant charge state.
\end{itemize}
It may be noted here that the gravitational effect is very important
but beyond the scope of our present work. In future and for
better understanding, someone can investigate the nonlinear
propagation in a three-component plasma by considering the
gravitational effect. Hopefully, we can emphasize that the results obtained from the investigation would be helpful to realize
various nonlinear phenomenon, where we can expect to have some possibilities for the occurrence of MI as well as the
construction of RWs in astrophysical ambience, space, viz.,
Jupiter's magnetosphere \cite{Horanyi1993,Horanyi1996},
cometary tails \cite{Horanyi1996}, Earth's mesosphere \cite{Havnes1996}, Saturn's rings \cite{Havnes1995}, and also
laboratory plasmas, viz., Q-machines \cite{Kim2006}, Coulomb-Crystal \cite{Chu1994,Thomas1994,Zheng1995}.

\end{document}